\documentclass[reprint,prl, groupedaddress, nobibnotes, nofootinbib]{revtex4-1}

\usepackage{graphicx}
\usepackage{dcolumn}
\usepackage{amsmath}
\usepackage{textcomp}

\begin{document}
\title{Enhancement-mode buried strained silicon channel quantum dot with tunable lateral geometry}
\author{T. M. Lu}
\altaffiliation{Contributed equally to this work.}
\affiliation{Sandia National Laboratories, Albuquerque, NM, 87123, USA}
\author{N. C. Bishop}
\altaffiliation{Contributed equally to this work.}
\affiliation{Sandia National Laboratories, Albuquerque, NM, 87123, USA}
\author{T. Pluym}
\affiliation{Sandia National Laboratories, Albuquerque, NM, 87123, USA}
\author{J. Means}
\affiliation{Sandia National Laboratories, Albuquerque, NM, 87123, USA}
\author{P. G. Kotula}
\affiliation{Sandia National Laboratories, Albuquerque, NM, 87123, USA}
\author{J. Cederberg}
\affiliation{Sandia National Laboratories, Albuquerque, NM, 87123, USA}
\author{L. A. Tracy}
\affiliation{Sandia National Laboratories, Albuquerque, NM, 87123, USA}
\author{J. Dominguez}
\affiliation{Sandia National Laboratories, Albuquerque, NM, 87123, USA}
\author{M. P. Lilly}
\affiliation{Sandia National Laboratories, Albuquerque, NM, 87123, USA}
\author{M. S. Carroll}
\affiliation{Sandia National Laboratories, Albuquerque, NM, 87123, USA}

\date{\today}

\begin{abstract}

We propose and demonstrate a relaxed-SiGe/strained-Si (SiGe/s-Si) enhancement-mode gate stack for quantum dots.  The enhancement-mode SiGe/s-Si structure is pursued because it spaces the quantum dot away from charge and spin defect rich dielectric interfaces and minimizes background dopants.  A mobility of $1.6\times10^5$ cm$^2$/Vs at $5.8\times10^{11}$/cm$^2$ is measured in Hall bars that witness the same device process flow as the quantum dot. Periodic Coulomb blockade (CB) is measured in a double-top-gated lateral quantum dot nanostructure. The CB terminates with open diamonds up to $\pm$10 mV of DC voltage across the device. The devices were fabricated within a 150 mm Si foundry setting that uses implanted ohmics and chemical-vapor-deposited dielectrics, in contrast to previously demonstrated enhancement-mode SiGe/s-Si structures made with AuSb alloyed ohmics and atomic-layer-deposited dielectric.  A modified implant, polysilicon formation and annealing conditions were utilized to minimize the thermal budget so that the buried s-Si layer would not be washed out by Ge/Si interdiffusion.

\end{abstract}

\maketitle

Control over single electrons, their spin and controlled coupling between electrons have reached exquisite levels in model semiconductor systems such as GaAs \cite{Petta2005, Koppens2006}.  The historical success of semiconductors for computation combined with demonstrations of the necessary electronic control of electron spins to do coherent manipulations of the spin state offers the promise that quantum dots might be used for quantum computation.

There continues to be considerable interest in translating the GaAs results into Si, because of the promise that very long spin decoherence times may be achieved\cite{Tyryshkin2003, Witzel2010}.  Significant progress in Si has recently been reported, particularly with single spin manipulation \cite{Hayes2009, Xiao2010, Morello2010, Simmons2011}. However, the long-term chance for success in Si, as well as near-term efforts to move towards controlling more spins coherently (e.g., double dots), will benefit greatly from both as smooth a disorder potential as possible, and minimal inhomogeneity in the local magnetic field. Common sources of unintentional charges and spins include interface states at semiconductor/dielectric interfaces and background dopants\cite{Nordberg2009a, Witzel2010}. Common choices for Si quantum dot structures either rely on metal-oxide-semiconductor enhancement-mode configurations \cite{Liu2008,  Nordberg2009b, Lim2009, Xiao2010} or SiGe/strained-Si (SiGe/s-Si) modulation-doped configurations \cite{Simmons2007, Hayes2009}, both of which intrinsically introduce charge and spin centers near the target spin either through interface defects or a necessary partially ionized supply layer. 

In this paper we propose and demonstrate an enhancement-mode SiGe/s-Si stack with a gate-defined quantum dot, which both spaces the quantum dot spin away from interface defects and removes dopants.  Enhancement-mode SiGe/s-Si field-effect transistors (FETs) have been demonstrated previously \cite{Klauk1996,Lu2007} and have recently achieved mobilities as high as $1.6\times10^6$ cm$^2$/Vs \cite{Lu2009}.  Enhancement-mode structures furthermore may provide increased tunability over critical properties such as valley splitting through tuning of the vertical electric field\cite{Friesen2007}. We report in this letter that low-disorder quantum dots can be formed using a double-top-gated lateral quantum dot nanostructure, defined using deep ultraviolet (DUV) 248 nm lithography and 150 mm-wafer processing.  We measure a Hall mobility of $1.6\times10^5$ cm$^2$/Vs at 4 K, showing that relatively high mobilities, indicative of low disorder, can be sustained in a foundry process using implanted ohmic contacts and chemical-vapor-deposited (CVD) dielectrics.

\begin{table*}

\caption{\label{fabprocesstable}Process Flow}
\begin{ruledtabular}
\begin{tabular}{c|c|c}
Sandia National Labs foundry process & Process Details & Thermal Budget\\
\hline
Si$_3$N$_4$ deposition & LPCVD & 750$^\circ$C, 85 min\\
Alignment marks; ohmic implant & $5\times10^{16}$/cm$^2$, 130 keV & \\
Amorphous Si deposition & LPCVD & 550$^\circ$C, 129 min\\
Poly implant; recrystallization anneal & $5\times10^{15}$/cm$^2$, 35 keV &700$^\circ$C, 30 min\\
Poly etch; poly reoxidation & & 750$^\circ$C, 46 min\\
RTA &&800$^\circ$C, 10 sec\\
Si$_3$N$_4$ deposition & PECVD & 400$^\circ$C, 47 sec\\
SiO$_2$ deposition & HDPCVD & 400$^\circ$C, 105 sec\\
CMP; via etch; Ti/TiN deposition; RTA &sputtering &750$^\circ$C, 30 sec\\
W/TiN deposition; metal etch; window etch; forming gas anneal &CVD&400$^\circ$C, 30 min\\

\end{tabular}
\end{ruledtabular}
\end{table*}

The starting material for our devices was a s-Si/relaxed-SiGe/s-Si/relaxed-SiGe epitaxial stack, grown on a SiGe virtual substrate. All epitaxy, done using low-pressure chemical-vapor-deposition (LPCVD) at Lawrence Semiconductor Research Laboratory, started on 5-30 $\Omega$cm, B-doped (100) Si substrates with $<$ 1.0 degree miscut.  The virtual substrate consisted of a 3 $\mu$m-thick linearly-graded layer and a 1 $\mu$m-thick relaxed SiGe layer, which was smoothed by chemical mechanical polishing (CMP).  The Ge concentrations of the SiGe layers and the virtual substrate were nominally 30\%. Following this, an additional SiGe buffer layer (200nm), a s-Si quantum well (15nm), a SiGe barrier (95nm), and a s-Si cap (2nm) were grown.  

In the Sandia National Labs silicon foundry, the device stack was formed with a 50 nm-thick LPCVD Si$_3$N$_4$ layer, a 200 nm-thick As-doped polysilicon layer, and an oxide/nitride/oxide stack capped with a Ti/TiN/W metallization layer. Ohmic contacts were formed with degenerately doped As implants. The polysilicon was patterned using lithography with approximately 180 nm resolution. The first oxide was formed by steam oxidation of the polysilicon after etching, and the nitride/oxide stack was deposited by plasma-enhanced CVD (PECVD) of 35 nm of Si$_3$N$_4$ followed by 800 nm of high-density-plasma CVD (HDPCVD) SiO$_2$. A CMP step was used to planarize the device structure before via etches to the ohmics and metallization was completed. The top metal stack, 20 nm Ti/50 nm TiN/100 nm W/25 nm TiN, was deposited after vias to the ohmic implant regions were etched. Table \ref{fabprocesstable} summarizes the process flow. A cross-sectional schematic of the devices is drawn in Fig.~1(a).

Strain relaxation and interdiffusion of Ge into the buried s-Si layer introduce limits on the thermal budget available for device processing.  High quality dielectric deposition and dopant activation often use high temperatures, well above the growth temperatures of the SiGe/s-Si stack.  Rapid thermal annealing (RTA) at 800$^\circ$C for 10 seconds activates enough dopants to avoid freeze-out down to the lowest temperature measured, 0.25 K. The resistivity of the implanted regions is 155 $\Omega$/square at 4 K. The thermal budget of the dielectric layer depositions and the activation anneal are indicated in table \ref{fabprocesstable}.  X-ray diffraction (XRD) of epitaxy on wafers grown under similar conditions as the device wafer indicated a starting s-Si quantum well thickness of 15$\pm$2 nm. Cross-sectional transmission electron microscopy (XTEM) of the processed device structure, displayed in Fig.~1(b), shows that the buried s-Si layer is approximately 6 nm thick clearly indicating that the s-Si well does survive the thermal budget.  The starting s-Si well thickness was not directly quantified by XTEM, because the entire 150 mm wafer was necessary for fabrication in the silicon foundry, and the final s-Si well thickness was not measured by XRD as it was below the detection limits of the measurement.

\begin{figure}[h]
\resizebox{3.5 in}{!}{\includegraphics{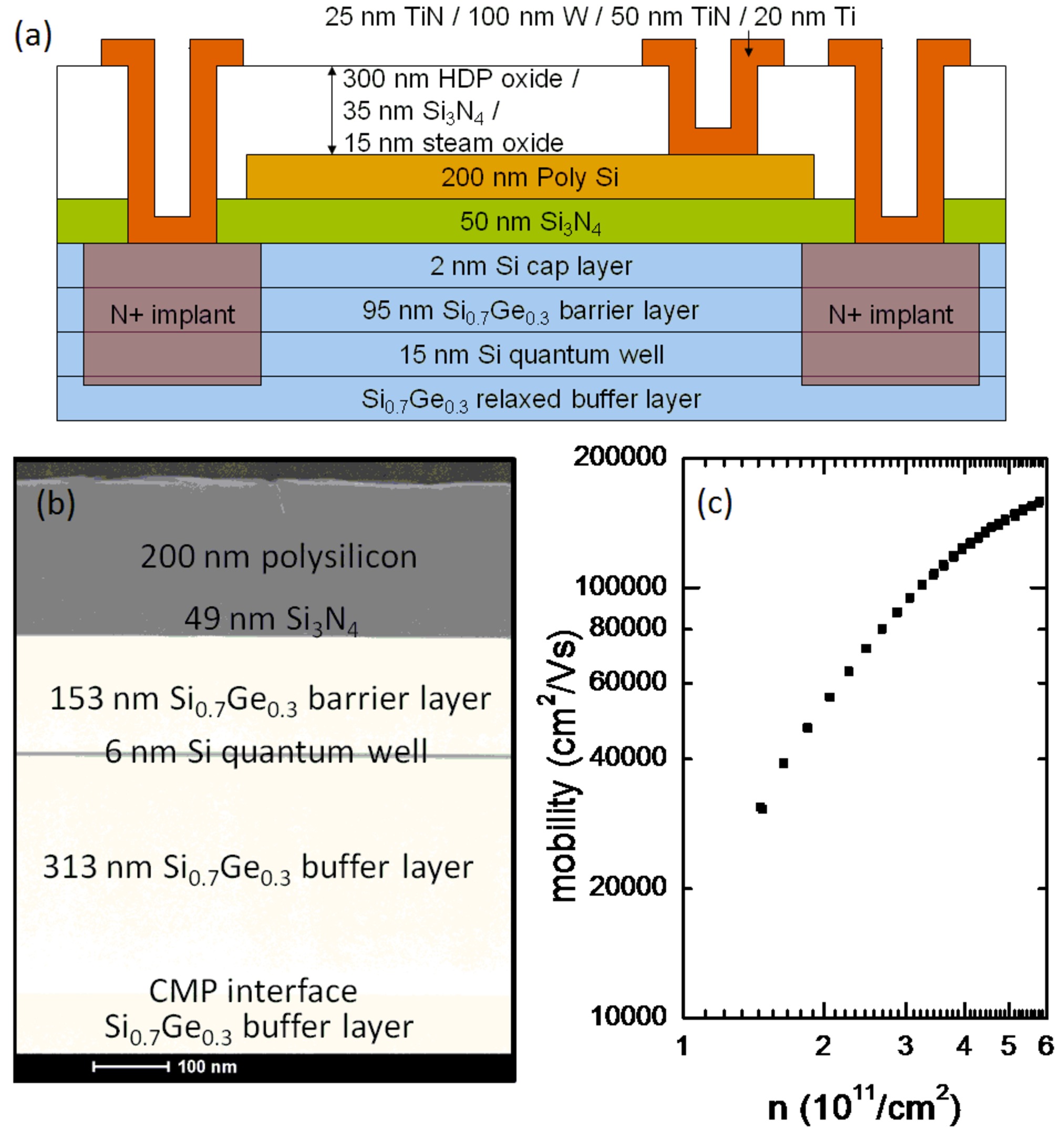}}
\caption{\label{fig1}(Color online) (a) Cross section of device used for Hall measurements. (b)  XTEM of post-processed device stack. A thin interfacial layer where the nominal 2 nm Si cap layer was grown is seen at higher magnification (not shown). (c) The electron mobility dependence on density obtained from standard low-field Hall measurements. The peak density is limited by onset of leakage in the device.}
\end{figure}

We measured the dependence of mobility on density using $100\times600~\mu$m$^2$ Hall bars.  The polysilicon layer was used as an enhancement gate in the Hall bar structure. Low-field Hall effect measurements were carried out at 4K using standard lock-in techniques, and the results are shown in Fig.~1(c). The mobility increases with density to $1.6\times10^5$ cm$^2$/Vs at $5.8\times10^{11}$/cm$^2$. Onset of gate leakage limits the maximum density.  We observed similar mobilities in a second wafer, identical to the first but for a nominally 80 nm-thick SiGe barrier layer. Ge/Si interdiffusion, expected with this thermal budget, will lead to a less abrupt interface and a thinner quantum well, both of which presumably reduce the mobility.  The highest reported mobility \cite{Lu2009} is for a case where the highest thermal step is kept at 440 C.  Smoother disorder potentials and higher mobilities are therefore possible if desired in this particular stack considering that several thermal steps in the process flow have not yet been optimized.

\begin{figure}[h]
\resizebox{3.5 in}{!}{\includegraphics{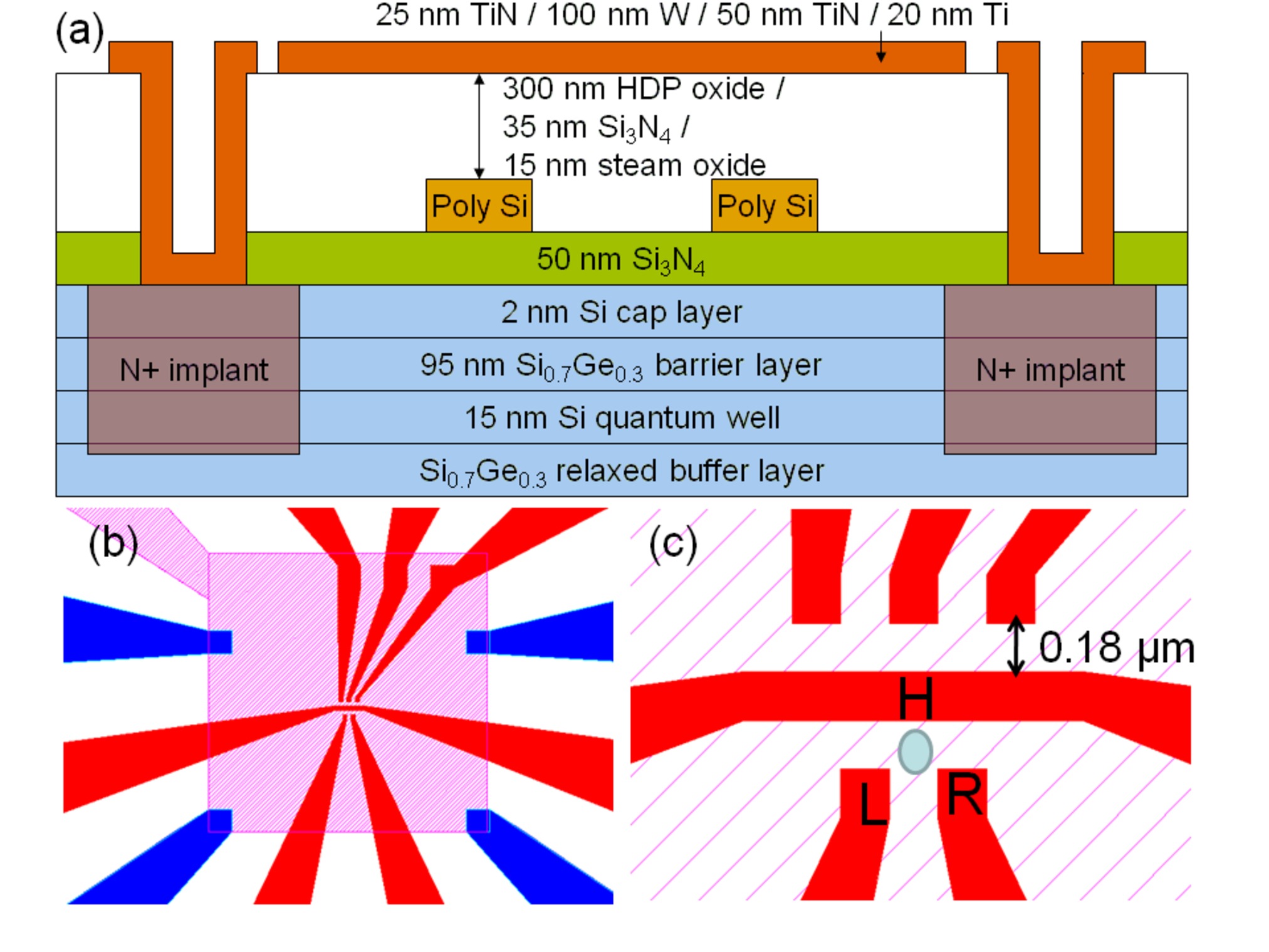}}
\caption{\label{fig2}(Color online) (a) Cross section of double-top-gated nanostructure. Topmost metal stack is used to induce electrons and the buried polysilicon gates locally deplete. No intentional doping is in this device stack. (b) Schematic layout of the supporting peripheral regions for the quantum dot. Blue indicates the location of degenerately As-doped layers, red indicates the location of the buried polysilicon layers, and hatched magenta indicates the overlaying metal gate. (c) Schematic layout of the quantum dot. Red indicates the location of the buried polysilicon gates. The ellipse is a cartoon representation of where the dot might be forming.}
\end{figure}

Nanostructures, fabricated on the same 150 mm wafer, were operated in a double-top-gated configuration. The top metal gate was positively biased to accumulate electrons in the s-Si quantum well globally.  In contrast with the Hall bar devices, here the nano-patterned polysilicon gates, below the top metal gate, were negatively biased to isolate the quantum dot region. A schematic drawing of a double-top-gated device is shown in Fig.~2(a). Figure 2(b) displays the design of the quantum dot studied in this work, and a zoom-in of the active region is provided in Fig.~2(c). In this letter, we present transport measurements through the bottom dot formed between the 'H', 'L', and 'R' depletion gates. 

\begin{figure}[h]
\resizebox{3.5 in}{!}{\includegraphics{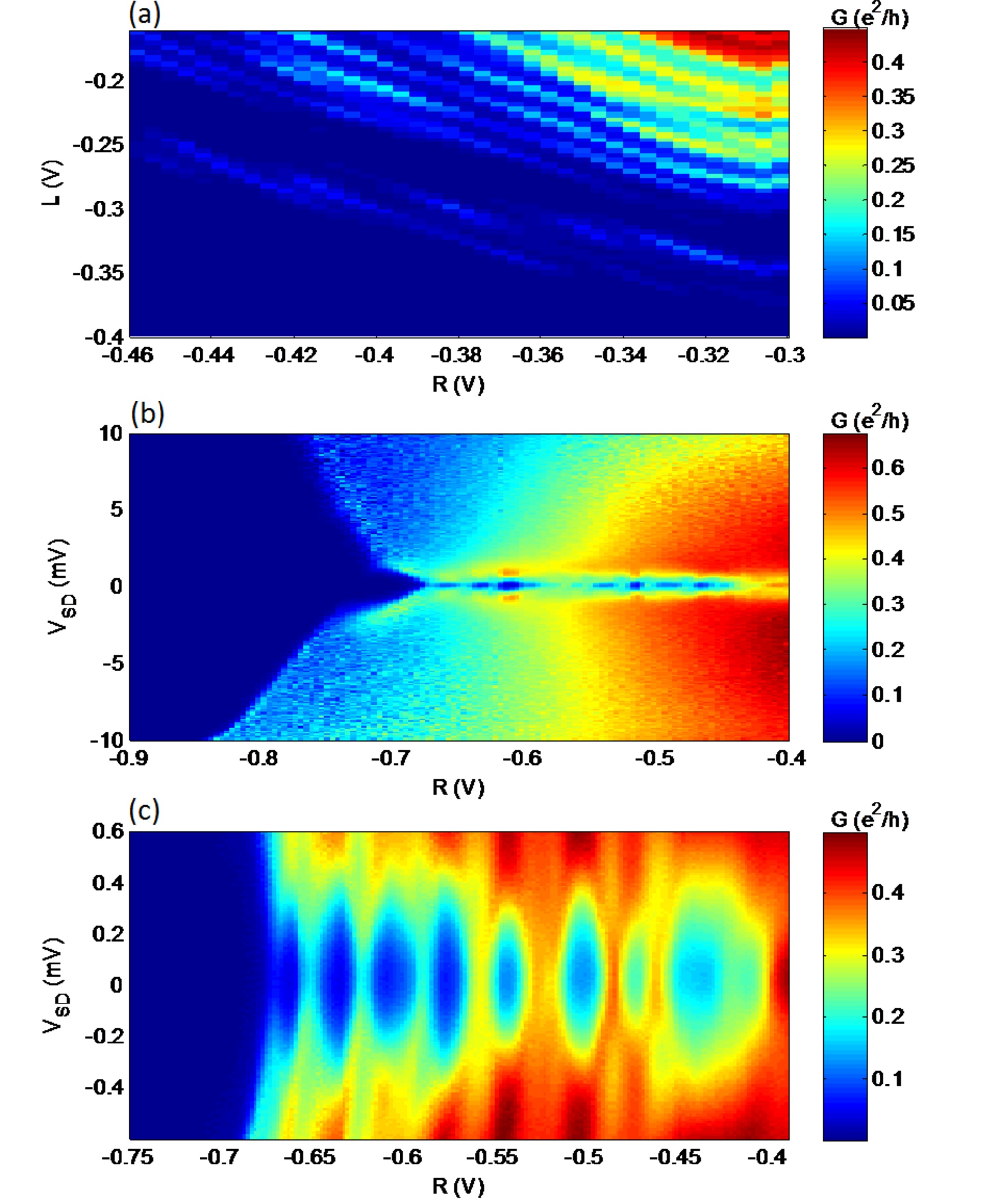}}
\caption{\label{fig3}(Color online) (a) Dependence of the nanostructure's conductance on gates L and R. V$_{top gate}=34$V and V$_H=-0.75V.$ (b) Conductance dependence on DC V$_{SD}$ and depletion gate R. V$_L=-0.3$V. (c) Finer resolution of nanostructure conductance dependence on V$_{SD}$ and V$_R$.}
\end{figure}

The transport experiment was performed in a $^3$He cryostat with a base temperature of 0.25 K using standard lock-in techniques. The excitation voltage across the dot was 10 $\mu$V at 79 Hz. Figure 3(a) shows the conductance with the 'L' and 'R' gates swept against one another. Periodic resonances are observed in the upper right quadrant of the plot, consistent with Coulomb blockade and quantum dot formation. The slope of the resonances indicates that the dot couples to the two depletion gates with roughly equal capacitance.  The slope of the resonances is uniform over the bias range, indicating a well-formed single-dot, without any nearby parasitic dots. To measure the charging energy, we apply a combination of DC and AC voltages on the drain, and measure the differential conductance, shown in Figs.~3(b) and (c).  Most of the Coulomb diamonds extend to approximately 300 $\mu$eV, though after the last transition observable at zero DC bias, no further transitions are observed up to DC voltages as high as $\pm$10 mV suggesting the dot has low electron occupation.

In summary, we propose and demonstrate a SiGe/s-Si enhancement-mode gate stack for quantum dots. This structure is appealing for quantum computing applications because it is a path towards minimizing background charge and spin defects relative to other Si quantum dot configurations. The devices are fabricated in a 150 mm silicon foundry requiring the use of implanted ohmics and CVD dielectrics, in contrast with previous demonstration of high-mobility SiGe/s-Si enhancement-mode FETs that used AuSb alloyed ohmics and low-temperature atomic-layer-deposition of Al$_2$O$_3$ for the gate dielectric. A mobility of  $1.6\times10^5$ cm$^2$/Vs at $5.8\times10^{11}$/cm$^2$ at 4K is established in Hall bars. Higher electron densities were limited by leakage in these device structures. Double-top-gated nanostructures using the same process flow show periodic Coulomb blockade. The quantum dot structure uses a remote global gate to induce electrons, polysilicon gates patterned with DUV lithography and a CVD Si$_3$N$_4$ layer to form the gate dielectric/semiconductor interface. The Coulomb blockade is consistent with a lithographically defined dot with charging energy around 300 $\mu$eV.

The authors would like to acknowledge useful discussions with Professor J. Hoyt (MIT) and K. Childs.   This work was performed, in part, at the Center for Integrated Nanotechnologies, a U.S. DOE, Office of Basic Energy Sciences user facility.   The work was supported by the Sandia National Laboratories Directed Research and Development Program.  Sandia National Laboratories is a multi-program laboratory operated by Sandia Corporation, a Lockheed-Martin Company, for the U. S. Department of Energy under Contract No. DE-AC04-94AL85000.

\end{document}